\documentclass[12pt,preprint]{aastex}

\begin{document}
\bibliographystyle{apj}

%  LaTeX definitions 
%
\def\slantfrac#1#2{\hbox{$\,^{#1}\!/_{#2}$}}
\def\figsize{\epsfxsize=1.0\columnwidth}
\def\ltsim{\raisebox{-.5ex}{$\;\stackrel{<}{\sim}\;$}}
\def\gtsim{\raisebox{-.5ex}{$\;\stackrel{>}{\sim}\;$}}
\def\hi{H\,{\sc i}}
\def\hii{H\,{\sc ii}}
\def\hei{He\,{\sc i}}
\def\heii{He\,{\sc ii}}
% end LaTeX definitions
%
\title{Grain Physics and Rosseland Mean Opacities}

\author{ Jason W. Ferguson\altaffilmark{1,2},
Amanda Heffner-Wong\altaffilmark{1,3},
Jonathan J. Penley\altaffilmark{1},
Travis S. Barman\altaffilmark{4},
David R. Alexander\altaffilmark{1,5}
}

\altaffiltext{1}{Department of Physics, Wichita State University, Wichita, KS 67260-0032; 
jason.ferguson@wichita.edu}
\altaffiltext{2}{visiting Max-Planck-Institut f\"{u}r Astrophysik, Karl-Schwarzschild-Str. 1, 85748
Garching, Federal Republic of Germany}
\altaffiltext{3}{now at Project Zero, Harvard Graduate School of Education, 
124 Mount Auburn Street, 5th Floor, Cambridge, MA 02138; amanda\_heffner-wong@pz.harvard.edu}
\altaffiltext{4}{Lowell Observatory, 1400 W. Mars Hill Rd., Flagstaff, AZ 86001; 
barman@lowell.edu}
\altaffiltext{5}{now at Idaho State University; alexdavi@isu.edu}

\begin{abstract}
Tables of mean opacities are often used to compute the transfer 
of radiation in a variety of astrophysical simulations from stellar
evolution models to proto-planetary disks.  Often tables, such as \cite{f05}, are computed with 
a predetermined set of physical assumptions that may or may not be valid for a specific application.
This paper explores the effects of several assumptions of grain physics on 
%### added phrase for clarity
the Rosseland mean opacity in an oxygen rich environment.  
We find that changing the distribution of grain sizes, either the power-law 
exponent or the shape of the distribution, has a marginal 
effect on the total mean opacity.  
%### rewrote sentence reflecting change in section 2.3
%A more important assumption is one of solid homogenous grains as 
%opposed to grains that are porous or conglomorations of several species.  
We also explore the difference in the mean opacity between solid homogenous grains
and grains that are porous or conglomorations of several species.  
Changing the amount of grain opacity included in the mean
by assuming a grain-to-gas ratio significantly affects the mean opacity, but in a predictable way.
\end{abstract}


% uncomment next line for preprint version
%\newpage

\section{Introduction}

The opacity of gaseous environments is a topic of interest in many areas of astrophysics
including stellar atmospheres, circumstellar and protoplanetary disks, and the interstellar
medium.  Often, at least in the 
case of stellar evolution calculations, opacity tables are needed at many temperature,
density and chemical composition points.  For high temperature opacities
modellers often use either
the OPAL opacities (\nocite{ir1991} \nocite{ir1993} \nocite{ir1996}
Iglesias \& Rogers (1991, 1993, 1996), \nocite{ri1992a} \nocite{ri1992b} 
Rogers \& Iglesias (1992a, 1992b) and \cite{rsi1996}) or the Opacity Project
\cite[OP hereafter]{OP1994} tables. 
We have produced low temperature opacities at Wichita
State University \cite[F05 hereafter]{f05} since 1975.  These tables differ from their
higher temperature counterparts by including the effects of molecules and dust in the
equation of state (EOS hereafter) and the opacity.

Typical opacity tables provided by our group include over
1600 temperature and density points with 155 chemical compositions needed for
a complete set.  Given the large number of input variables, atomic and molecular physics, grain
parameters for grain physics, etc., it is not computationally possible to make 
a complete set of opacity tables for every astrophyisically significant situation.  Difficult 
decisions must be made as to how to choose important physical parameters.  For example,
in the opacity tables of F05 grains are assumed to have a size distribution 
like the one described in a classic model 
\cite[MRN hereafter]{mrn} based upon grain sizes in the interstellar medium (ISM);
however this choice may not be appropriate for
different astrophysical environments.  This paper discusses some of the choices made in F05
and how those choices affect the total Rosseland mean opacity.

The calculations discussed here are taken from the same code used to complete the 
opacity tables
described in F05, which includes a summary of the equation of state and opacities.  
For consistency we begin with the total Rosseland mean opacity, which is defined as

\begin {equation}
  \frac{1}{\kappa_R} \equiv \frac{ \int_0^\infty \frac{1}{\kappa_{\lambda}} \frac{\partial B_{\lambda}}{\partial T} d\lambda }
     {\int_0^\infty \frac{\partial B_{\lambda}}{\partial T} d\lambda }
\end {equation}

\noindent 
where $\kappa_{\lambda}$ is the monochromatic opacity and 
$\partial B_{\lambda} / \partial T$ is the derivative of the Planck function with respect to
temperature.  The monochromatic opacity includes the effects of all contributors:

\begin {equation}
  \kappa_{\lambda} = \kappa_{cont} + \kappa_{at} + \kappa_{mol} + \kappa_{gr} 
\end {equation}

\noindent 
and includes continuous, atomic, molecular and grain sources of opacity.  Note that the 
Rosseland mean is a harmonic mean, one in which individual contributors cannot be added after
the mean is taken.  This is particularly true of the molecular opacities which may 
have a large number of individual lines or bands.  If it is desired that an individual source of opacity 
be modified, it is usually the case that the whole integral must be recomputed.
A detailed description of these sources is given in F05.  We focus in this 
paper on the details of the grain contribution to the monochromatic opacity.  

The opacity due to a particular dust species can be computed from

\begin {equation}
\kappa_{gr} ~\rho~=~\pi\sum_{i}\int_a n_i(a)~Q_{\rm ext}(a,i,\lambda)a^2~da
\end {equation}

\noindent 
%### fixed units of n_i(a)...
where $n_i(a)$ is the number of dust particles (cm$^{-3}$) of species {\it i} of size {\it a} 
and $Q_{\rm ext}(a,i,\lambda)$ is the total extinction (absorption plus scattering) 
efficiency of the particle.  The size distribution $n_i(a)$ depends upon both the number 
abundance of species {\it i} and on the size distribution of the dust particles.  
Ordinarily in our opacity tables the size distribution is taken from MRN, 
the classic distribution of dust sizes in the ISM given by 

\begin {equation}
  n_i(a) = ka^{q}
\end{equation}

\noindent 
where {\it k} is a normalization constant of the distribution and $q$ is the exponent 
of the power law.  In F05 the value of $q$ is taken to be the classical MRN value of -3.5.  

Additional work by many other groups have resulted in other ``favored'' size distributions, however
%### change of language for referee
none have fully superseded the work of MRN (\cite{clayton2003}).  Other distributions such as
Kim et~al. (1995, KHM\nocite{kmh} hereafter) and Weingarnter \& Draine (2001, WD\nocite{wd} hereafter)
use power-laws as their 
basis, but include transitions for large grains so that the distribution behaves smoothly.  
%### mistake in references...
For carbon grains WD, using the work of \cite{lidraine1} and \cite{lidraine2} 
go a step further and include multiple distributions for different sizes of grains.
While this is not an 
exhaustive list of grain size distributions used in the astrophysical community 
today, it is representative of the types of grain models available.  Each of the distributions 
listed have been inserted into our code for comparison with the MRN distribution. 

In Eq.~3, the grain extinction efficiency is computed with Mie theory and assumes that grains are solid.
This may not be the case in all, or any, astrophysical environments.  
Many researchers argue that interstellar
%### added references for referee...
grains are likely to be porous or ``fluffy'' in nature (see for example \cite{mw1989}, \cite{mathis1996}, \cite{chiar06}).
%### cleaned up grammar and long sentence
Having grains which are part vacuum can significantly affect
their opacity (see \cite{wolff94}, \cite{wolff98} and \cite{VIH2005}).  
However, in astrophysical grains the amount of porosity, generally defined as the volume fraction
of vacuum, is not well constrained; \cite{chiar06} interpret observations to constrain 
%### new phrase as suggested by referee
the porosity of dust in the ISM to be in the 25\% to 50\% range whereas \cite{mathis1996} indicates
that dust mayby up to 80\% vacuum.

Related to the porosity of grains is whether or not grains are composed of aggregates or
are homogeneous.  It is likely that astrophysical grains are not homogeneous as is assumed in 
the opacity tables of F05, but rather are an aggregate of two, three or more different species.  
The effect of heterogeneous grains on the mean opacity using the utilities of 
\cite{Ossenkopf91} will be discussed below.

The opacity tables of F05 assume that gas and dust are in equilibrium, but this may
not be the case in real environments due to dynamical effects such as winds or gravitational settling 
of grains.  One simple way to mimic the effects of grains either settling or being blown out of
an environment is by changing the grain-to-gas ratio by assuming 
certain percentages of grain material have been removed from the gas.

This work is an exploration of the effects of grain physics on the total 
Rosseland mean opacity of a gas at cool temperatures (below $\sim$3200~K).  Opacities
calculated here are 
%### define X and Z and add clarity for oxygen-rich.
for solar abundances, from \cite{gn93} scaled to X=0.7 and Z=0.02 (where X is the mass fraction of hydrogen and 
Z is the mass fraction of metals), for 2.70 $\leq~\log T~\leq$ 3.30
and constant $\log R = -3.0$ (see F05).  The value of $\log R = -3.0$ best models the 
atmospere of an AGB-type star and was arbitraily chosen here as a convenient
%### oxygen-rich...
way to demonstrate grain physics effects.  All comparisons are done
in an oxygen rich chemistry and include only silicate grains, future work will emphasize
carbon grains.  We will discuss five important aspects of grains physics
in the following sections: (1) the size distribution exponent (see Eq. 4), (2) the size distribution
itself, (3) grain porosity, (4) aggregate grains, (5) and the relative grain-to-gas ratio.  

\section{Grain Physics}

In gas at high temperature, above $\sim$~5000~K, neutral and ionized atoms 
(including lines) are the dominant opacity sources.  As the temperature decreases, molecular
effects become significant and begin to dominate the opacity (see F05).  Below $\sim$1600~K
at $\log R=-3.0$ grains begin to appear in the equation of state and quickly become the strongest 
source of opacity.  For purposes of illustration, a plot similar to one from F05
is given in Figure~1 with the regions of dominant opacity marked.  The features 
below $\log T = 3.2$ are caused by various grain species appearing or disappearing from the 
equation of state.
At $\log T=3.2$, solid $Al_2O_3$ appears in the EOS and causes a sudden rise in total opacity 
as the temperature decreases.  Silicate grains begin to appear at $\log~T \sim3.08$.  
The opacity has another bump at $\log~T=3.02$ as
solid Fe appears and becomes the dominant source of opacity.  
At temperatures below $\log~T=3.0$ iron and silicates are the 
predominate opacity sources (F05); no additional 
significant species appear or disappear and the mean opacity is fairly featureless 
as it gradually lessens as the temperature decreases.

\begin{figure}
%\plotone{ahw_opac_plot.ps}
\plotone{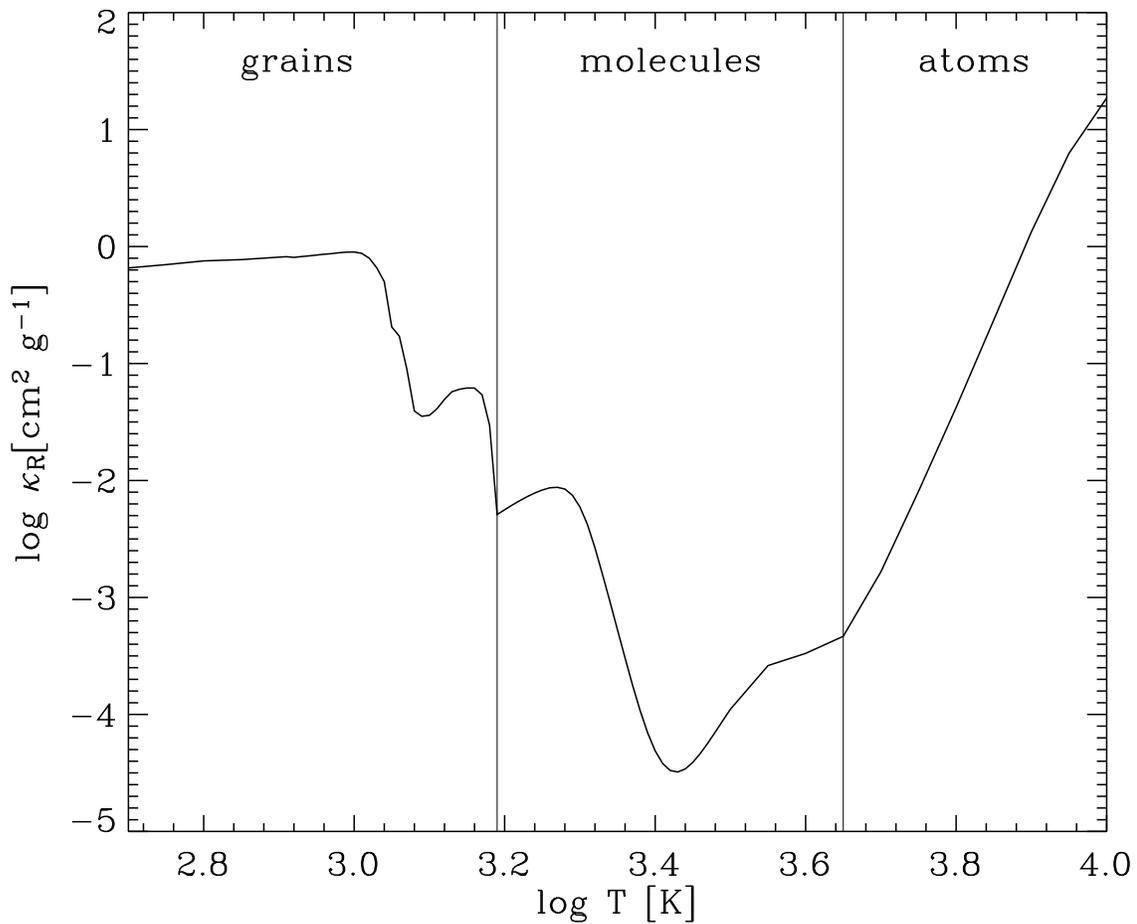}
\caption{Logarithm of the total Rosseland mean opacity as a function of temperature
for solar abundances and $\log R = -3.0$.  Regions where different components
dominate the opacity are indicated.  Low temperature features are discussed in the text.}
\label{Figure 1}
\end{figure}

The opacity tables computed in F05 assumed spherical, solid, homogeneous grains with sizes
that obey the MRN size distribution and with
absorption and scattering cross-sections that can be simulated with Mie theory.  We further
assume in F05 that the grains form in equilibrium with and remain with the gas they form in.  
We will relax a few of these assumptions in the following sections to explore 
how the Rosseland mean opacity is affected.

\subsection{Size Distribution Exponent}

Equation~4 is a normalized power law distribution.  
As the value of the exponent, $q$, is varied -- as the 
power law steepens or flattens -- the size distribution will vary.  As $q$ 
increases the function will become steeper, that is, there are more small
grains than large grains.  The distribution is normalized for each grain species so that
the volume of grain material is constant for the different values of {\it q}.  The 
normalizations and summations are performed in terms of $\log(a)$, where {\it a} is the 
grain size.  

As an illustration we include a plot of the ``guts'' of Eq.~3 in Figure~2.  The figure
shows how $Q(a)n(a)a^2$ behaves for a single grain species ($Al_2O_3$) and a single wavelength 
(1 $\mu$m).  As the exponent of the size distribution increases from -2.5 to -5.5, the
size distribution, $n(a)$, steepens, but $Q(a)n(a)a^2$ flattens out.  This flattening is due to the 
%### new phrase for clarity...
combination of the $a^2$ factor and the extinction efficiency $Q$ increasing with size 
(as the grains are smaller than the Rayleigh limit) for the species shown.

\begin{figure}
%\plotone{qext.n.a2.ps}
\plotone{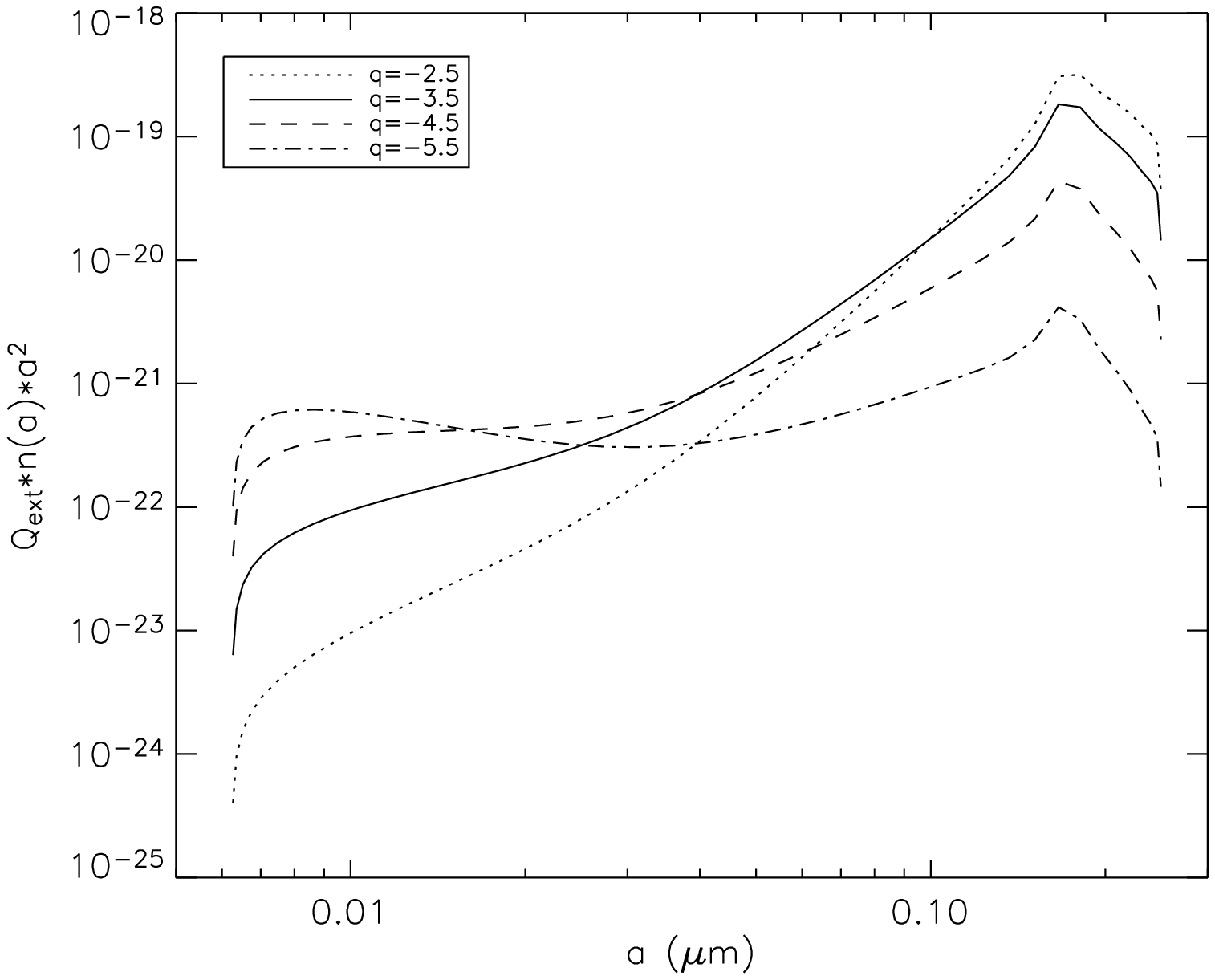}
\caption{Plot of $Q(a)n(a)a^2$ from Eq.~3 for $Al_2O_3$ at a single wavelength 
(1 $\mu$m).  The values shown for the power-law exponent are -2.5 (dotted), -3.5 (solid),
-4.5 (dashed), and -5.5 (dash-dot).}
\label{Figure 2}
\end{figure}

What happens to the mean opacity if the value of the power law exponent, $q$, 
is varied from $-2.5$ to $-5.5$?  Figure~3 shows that the mean opacity 
in the dust region ($\log T < 3.2$) increases as the power law flattens out.  
Considering Fig.~2 this result is not surprising.  As the 
grain size distribution, $q$, flattens out the extinction efficiency, which is a function
of grain size but not distribution, rises for larger grains.  Thus a flatter grain
size distribution contains more large grains and those larger grains contribute more 
to the total opacity.  

\begin{figure}
%\plotone{opac_comp_qext.ps}
\plotone{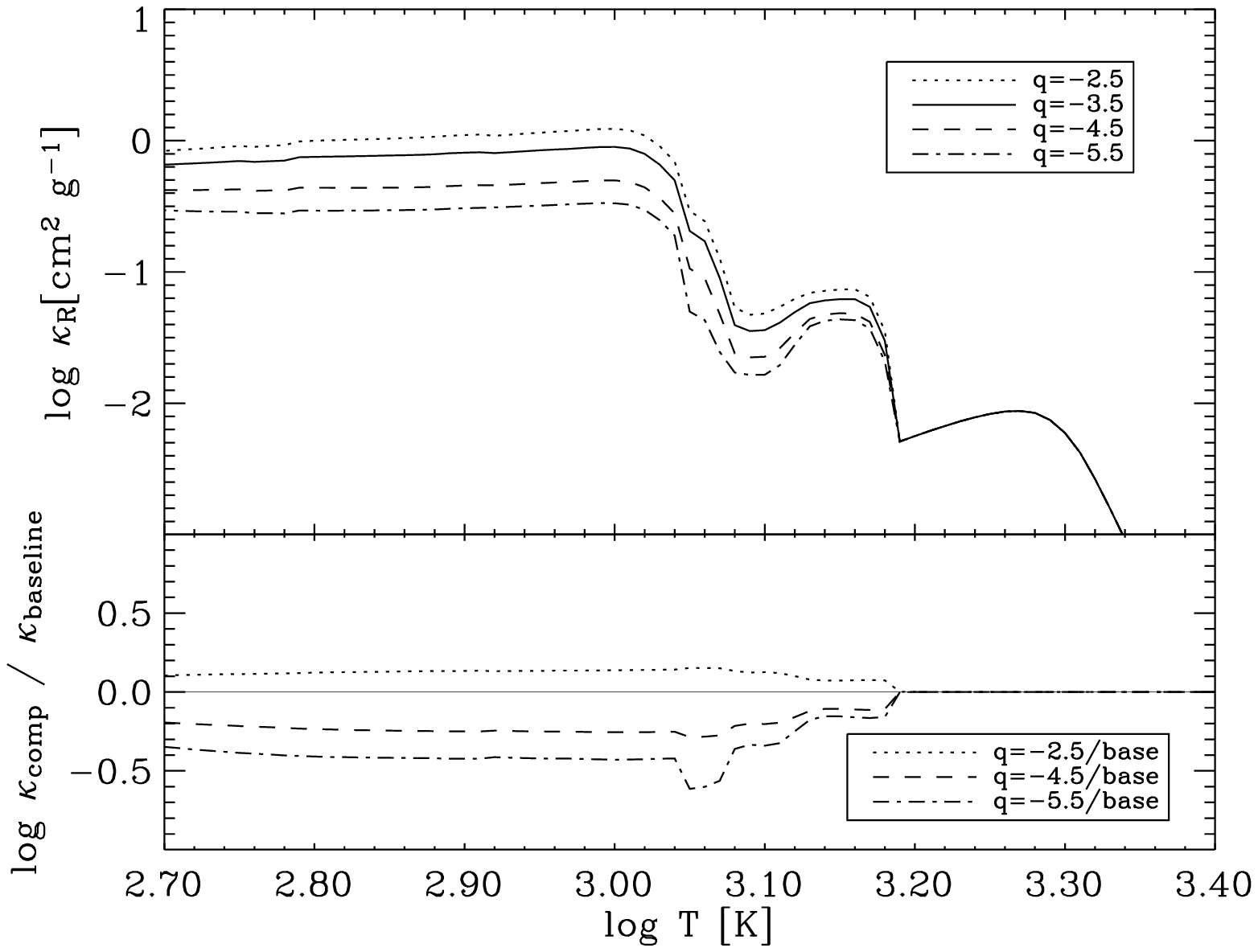}
\caption{Comparison of opacities calculated with different values of the MRN power law exponent, 
{\it q}, describing the power law size distribution of grains.  
In the upper panel values of {\it q} are, from top to bottom, $-2.5$ (dotted), $-3.5$ (solid),
$-4.5$ (dashed), and $-5.5$ (dash-dot). 
In the lower panel is the differences in the logarithm of the mean opacities for the various
test cases and the baseline ($q=-3.5$) are shown.  The dotted line is the logarithm of 
the ratio of the $q=-2.5$ case and the baseline, the dashed line is $q=-4.5$ case, and the dash-dot 
line represents the $q=-5.5$ case.  The thin solid line is included to help guide the eye
and is the value of the logarithm of one. 
}
\label{Figure 3}
\end{figure}

The overall result for the mean opacity is that a flatter grain size distribution
results in more opacity by as much as a factor of three when the power law exponent
changes from $-4.5$ to $-2.5$.  It is easy to see the difference changing the 
exponent, $q$, to the mean opacity with the lower panel of Fig.~3.  In the lower
panel the logrithm of the ratio of the mean opacity computed with a certain $q$ value 
compared with the baseline value of $-3.5$ is shown.  This type of plot
easily allows for the change in the mean opacity to be seen.

There are several features seen in the lower panel of Fig.~3 that deserve discussion
such as the stair step behavior of the comparison with the $q=-5.5$ computation with 
the baseline.  We include in our computations several species of grains in our total mean
opacity with their different opatical constants and extinction efficiencies.
As different species appear or disappear from the EOS the mean opacity is 
affected in a variety of ways based upon the dominant species.
More particularly, there is a large dip in the comparison with the $q=-5.5$ mean opacity 
between $3.04 \leq \log T \leq 3.08$.  
This feature, and the much smaller one seen in the $q=-4.5$ computation,
is due to the differing effects on changing the size distribution for different species.
In the lower panel of Fig.~3, at $\log T = 3.08$ the difference between the baseline
and the $q=-5.5$ computation begins to move downwards as the Mg-silicates begin to  
dominant the opacity.  The difference then moves upwards at $\log T = 3.04$ as
solid Fe becomes the dominant source of opacity (see F05).  Changing the size distibution 
in the same way for every species in our EOS has a different effect on that species'
contribution to the total mean opacity.  

Notice also that the comparisons with the $q=-2.5$ and $q=-4.5$ are not symmetrical
about the reference line 0.0.  This shows that changing the value of the exponent 
does not lead to a linear, or clearly defined change in the mean opacity
This fact is not surprising by noting in Fig.~2 that 
the extinction efficiency, a function of grain size, rises with $a$, but not in a ``linear" way.

It would be convienient to have a scaling factor to convert the baseline
opacities to a set of mean opacities with a different grain size distribution and
this may be possible away from transistion zones.  However, at those temperatures
where the dominate grain species are changing a simple scaling factor is not possible.

\subsection{Grain Size Distributions}
The size distributions discussed in Section~1 have been added to our opacity code and 
to be complete we include a short summary of the KMH and WD grain size distributions.
The size distribution from KMH is given by

\begin{equation}
  n_i(a) = k a^{q} exp(-a/a_b)
\end{equation}

\noindent where {\it a$_b$} is the transition size for the exponential decay.  
The values chosen for our computations are discussed below.

A more sophisticated grain distribution, 
which includes a smoother transition from the power-law to the expontial 
decay is given by WD for silicate dust and has the form

%$$
%1_Q(t) =
%\cases{
%   1, & if $t \in Q$\cr
%   0, & if $t \notin Q$\cr
%}
%$$

$$
\frac{dn}{da} = \frac{k}{a} \left( \frac{a}{a_{t}} \right)^{q} F(a;\beta,a_{t}) \\
$$
$$
\times 
\cases{1,                         & 3.5~\AA $< a < a_{t}$ \cr
       exp(-[(a - a_t)/a_{c}]^3), & $a > a_{t}$ \cr         }
$$

\noindent where {\it k} is a normalization constant, {\it q} is the power-law
exponent (originally $\alpha$ in WD), {\it a$_t$} and {\it a$_c$} provide
for the cutoff and shape of the distribution, and the power-law exponent represented by $q$.
The curvature of the function (changed by modifying $\beta$) is given by the term

$$
F(a;\beta,a_t) \equiv 
\cases{
  1+\beta a/a_t,        & $\beta \geq 0$ \cr 
  (1-\beta a/a_t)^{-1}, & $\beta < 0$ \cr
}
$$

Similar distributions are assumed by WD for carbon or graphitic grains which
include a log-normal distribution for the grain sizes to account for the very small
%### add phrase for referee.
grains thought emit the ``Unidentified Infrared'' (UIR) bands seen in the diffuse ISM.  The comparisons
in the present paper focus on oxygen-rich mixtures where carbon grains do not exist under
these conditions.  

For our comparisons we use the values given by WD for silicate dust in the ISM,
that is {\it a$_t$=0.17~\micron} and {\it a$_c$=0.1~\micron} 
with a value of $\beta$=0.3 and a power-law exponent of -2.1 for all grain types.  
The computations using the KMH
distribution use their stated parameters as well.  That is, {\it a$_b$} is taken to
be 0.14~{\micron} and a value of -3.06 for {\it q} is used for silicate dust.  

Figure~4 shows the comparsion between Rosseland mean opacities computed with 
grain size distributions from MRN, WD and KMH.  In the figure the 
mean opacites are shown in the upper panel and the differences in the 
logarithm of the opacities is shown in the lower panel.  
The results show that the mean opacities
are somewhat similar, with MRN having the lowest values, WD having higher values,
and KMH having the highest values of mean opacity.  It is 
interesting to note that KMH, with a very simple exponential
transition for large grains, tend to tail upwards at the lowest temperatures. 
At $\log T = 2.7$ (500~K), silicate grains are the dominant source of grain opacity, with
solid iron grains falling in abundance in favor of FeS (see Fig.~1 of F05).

There is also a strange series of bumps in the lower panel of Fig.~4
at intermediate temperatures, $3.00 < \log T <3.15 $.  
As discussed in the previous section, these temperatures are a transition
region where the strongest opacity source is passing from minor grain species to 
the silicates to solid Fe.  
Different grain species have different contributions 
to the opacity based upon which form of the size distribution is used due to different 
optical constants and extinction efficiencies.
The feature seen in the lower panel of Fig.~4, centered at $\log T = 3.08$, in the 
computations with KMH compared
with the baseline is much more pronounced than the computations with WD.
The difference between the KMH and WD grain distributions is how the size transition
to large grains is treated.  The minor grain species are much more
sensitve to the KHM distribution than the solid Fe grains are, since when
solid Fe dominates the opacity both the KMH and WD computations appear to be a
simple scaling factor greater in the mean opacity.

As discussed in the previous section, there does appear to be a simple scaling 
factor that might apply to the grain distibutions, but only for a limited range in 
temperature, $2.80 \leq \log T \leq 3.02$.  At other temperatures, where the 
strongest opacity source is changing from one grain species to another, no such
simple scaling factor is evident.

\begin{figure}
%\plotone{opac_comp_grndist.ps}
\plotone{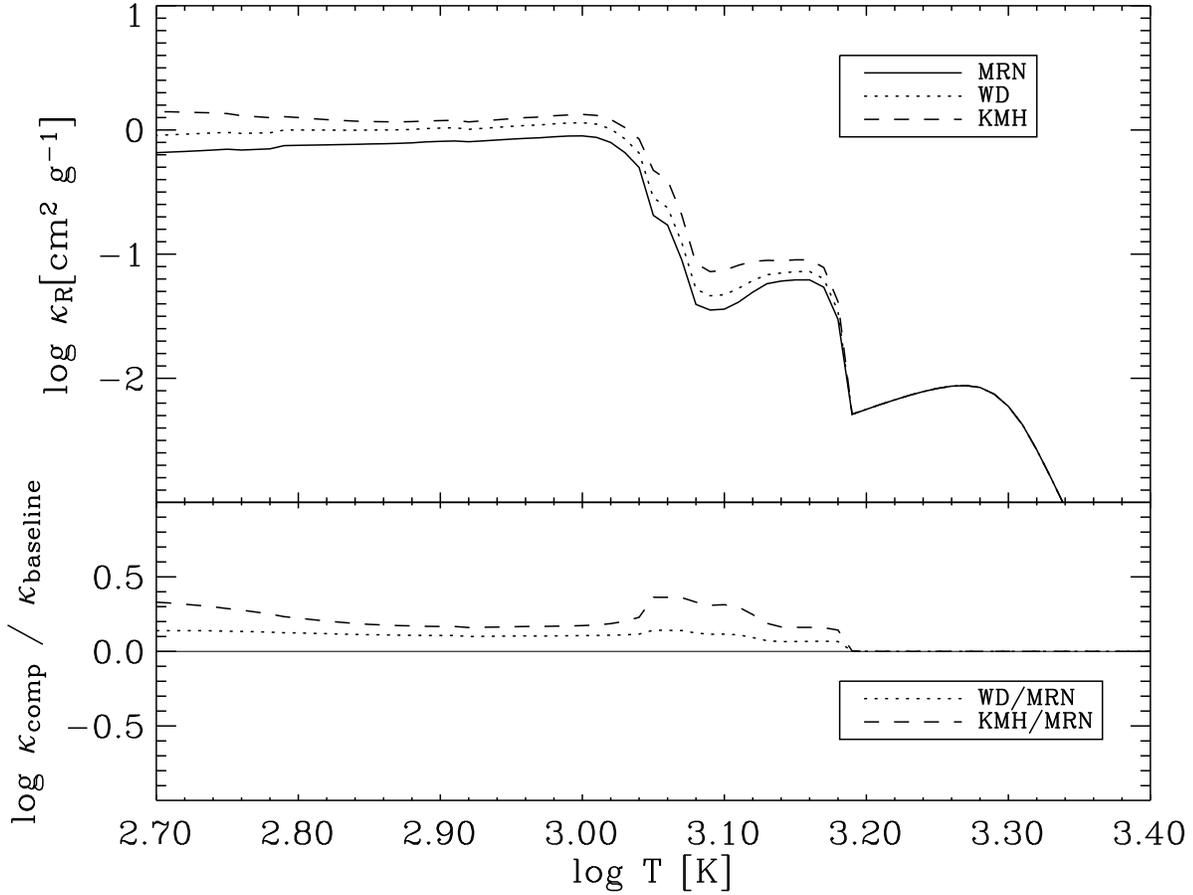}
\caption{Comparison of the mean opacity due to differing grain size distrbution models.  
In the upper panel, the solid line is the mean opacity when using the MRN distribution, 
WD is represented by the dotted line, and the model of KMH is shown by the dashed line.  
In the lower panel the differences are shown as in Fig.~3, with the thin solid line
at the 0.0 axis shows no change to the opacity, the dotted line compares WD with the baseline,and the dashed line compares KMH with the baseline.
See the text for detailed discussion.}
\label{Figure 4}
\end{figure}

Overall the effect on the mean opacity is fairly significant with the KMH calculation 
differing from the baseline (MRN) by 0.27 dex at $\log T = 2.9$ (800~K), or nearly a factor of 3.  As 
Fig.~4 shows the difference from the baseline is not constant with temperature, 
especially in the regions where the significant contributors to the opacity are from 
different grain species 
(with different extinction efficiences).  For example around $\log T\sim3.15$ the 
differences between the computations are smaller than at cooler temperatures.  
This implies that
arbitrarily adding a factor of 3 to the mean opacity tables of F05 to mimic the 
effects of different size distributions is not a good approximation.  This is particularly
true when combined with the mean opacity changes with differing size distribution exponents
as described above.

\subsection{Grain Porosity}
%### entire section rewritten as suggested by the referee.....
Calculations of grain efficiencies due to grain porosity are 
taken from \cite{bh83}.  To initially probe the effect of porosity on the 
mean opacity we assume that the vacuum inclusions are spherical and we 
use either the Maxwell-Garnett or the Bruggeman average dielectric constant 
for the grain/vacuum composite.  Maxwell-Garnett is more appropriate for a mixture
where the matrix and inclusions are well defined.  The Bruggeman approximation 
is more symmetric and valid for a two-component mixture with matrix and 
inclusions indistingishable.  As \cite{bh83} discuss both Maxwell-Garnett and Bruggeman
are equally valid, but either one or the other will be more valid in a particular circumstance.

In our computations we find that certain grain species
misbehave when using the Bruggeman approximation.  Most particularly troubling is
solid Fe.  At certain wavelengths the {\em k} portion of the optical constant will
be unphysically negative, causing the mean opacity to increase discontinuously by large 
factors.  
To test our method we have utilized the publically available average dielectric routines 
of \cite{Ossenkopf91} and found that program fails to compute porous Fe grains without either
blowing up with a fatal error, or producing negative parts of the optical constants.  We 
have decided to remove Fe from the porosity computation and let it remain solid and not porous.
We assume that the other grain species are valid under the Bruggeman assumption, the
EOS is mostly dominated by MgSiO$_3$ and Mg$_2$SiO$_4$ at 1000~K and below 
(see F05, Fig.~1).

Results for a range of grain porosities, denoted by the volume fraction $f$, are shown 
in Figure~5 with the mean opacities shown in the upper panel.  
The baseline value is $f$=0.0 (solid grains) and the tested values include
$f=$ 0.1, 0.25, 0.5, and 0.8.  Recent work by \cite{chiar06} report porosities of 
grains in the ISM as being between 25\% and 50\% vacuum by volume, 
in line with previous studies by \cite{wolff93}.
The lower panel shows the opacity differences compared with the baseline of
{\it f~=~0}.

\begin{figure}
%\plotone{opac_comp_porosity.ps}
\plotone{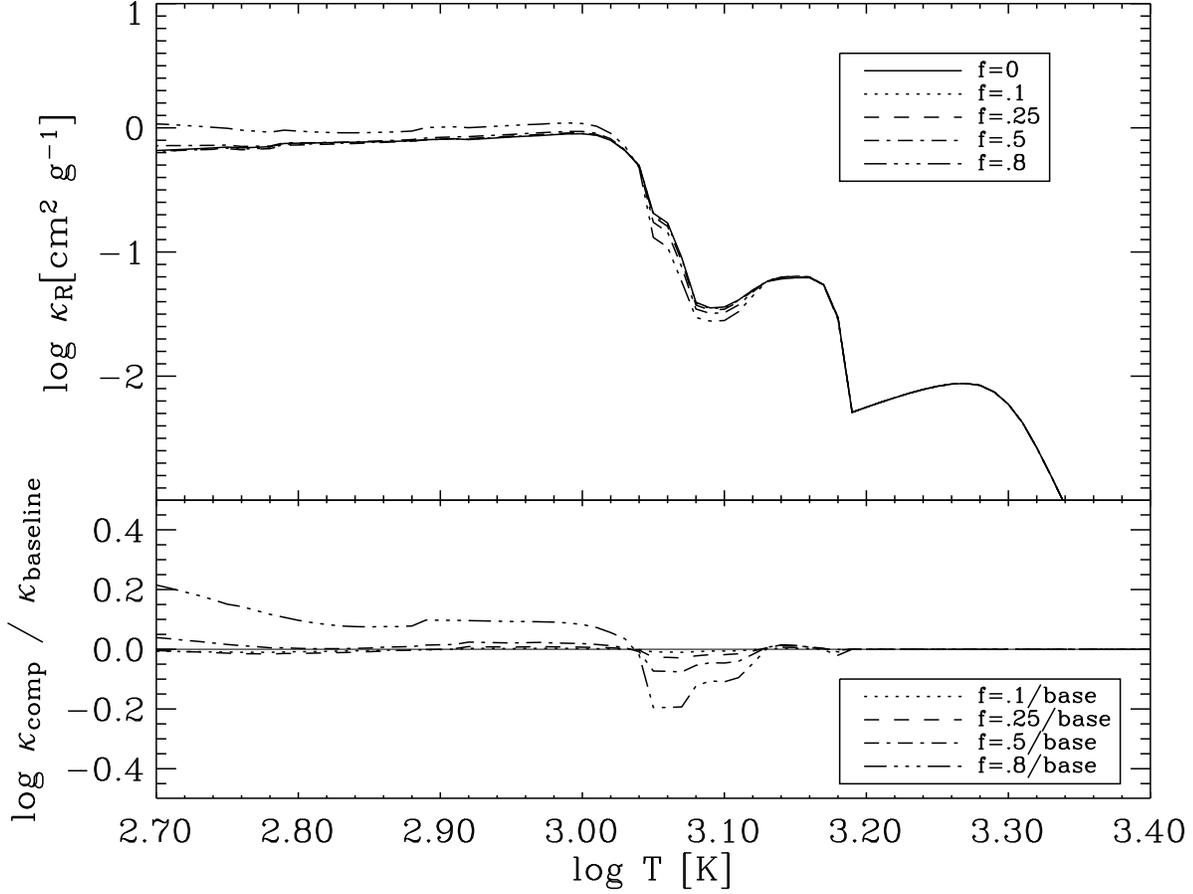}
\caption{Comparison of varying the porosity of grains.  In the upper panel, the 
mean opacities are shown.  The baseline calculation
is the solid line with dust porosity, {\it f}=0.0, the dotted line is {\it f}=0.1, 
the dashed line is {\it f}=0.35, the dash-dot line is {\it f}=0.5, and the 
dash-dot-dot line is {\it f}=0.8.
In the lower panel the differences are shown as in Fig.~3, with the porosity 
{\it f}=0.1 compared with the baseline as dotted, {\it f}=0.25 as dashed, {\it f}=0.5
as dash-dot, and {\it f}=0.8 as the dash-dot-dot line.
}
\label{Figure 5}
\end{figure}

As the porosity of grains is increased, the the total mean opacity 
is mostly unaffected, except for the largest $f$ value.  This is somewhat 
intuitive in that the mean opacity is already dominated by small particles, 
making those small particles slightly larger, does not change the mean opacity greatly.
At the largest value of porosity, $f=0.8$ the mean opacity does change by as much as 
50\% at the lowest temperature shown.  Note at ``intermediate'' temperatures, $3.05 < \log T < 3.13$,
the lower panel of Fig.~5 clearly shows that the mean opacity decreases 
with increasing porosity.  The materials responsible for
most of the mean opacity at these temperatures are SiO$_2$, 
calcium-silicates, and MgAl$_2$O$_4$ for the most part (see F05, Fig.~1).  
As these types of grains become larger and more porous that they also become
more transparent, unlike the silicates at cooler temperatures, where the mean opacity slightly increases
with porosity.

\subsection{Aggregrate Grains}
To explore the use of aggregate grains and their effects on the Rosseland mean opacity
we compute the average dielectric function as outlined by \cite{bh83} using
effective medium theory (EMT).  More specifically we assume 
Maxwell-Garnett theory for aggregates which is valid as long as the 
inclusions obey the Rayleigh limit as 
illustrated by \cite{wolff98}.  We use this simple EMT approximation rather than a more formal 
solution such as a 
discrete-dipole approximation (DDA) due to limits on computational time.  In addition, many 
of the features of a DDA are not needed for mean opacities computations, such as 
the scattering and polarizaton phase functions.

To compare with the baseline runs as outlined above we assume an aggregate grain that
is similar in composition to the silicate species computed in our EOS at $\log T = 2.9$ (800~K) and with 
ratios similar to silicates found in the inclusions of a typical meterorite such
as Semarkona as discussed in \cite{hewins96}.  While our EOS did not identically produce 
the Semarkona results, qualitatively it was fairly close; we compared our EOS with Semarkona 
and choose a median value.
The ratios of enstatite, forsterite, fayalite, and silicon dioxide were assumed to be
42\%, 36\%, 12\%, and 10\% respectively.  We computed the average dielectric functions using
the routines of \cite{Ossenkopf91} with the stated ratios.
Solid iron grains were kept separate from the aggregate as is found 
in meteoritic inclusions.  The mean opacity was then computed at a single temperature and
density ($\log R$) point and compared with the baseline.  Repeating this numerical experiment
at lower temperatures did not dramatically change the results

The comparisons show that aggregate grains have a lower mean opacity, 
on the order of 0.05 dex (12\%) below the baseline, which is a smaller change shown in 
Fig.~3 for changes in the grain size exponent $q$ from -3.5 to -4.5.  
The amount of change in the mean 
opacity for aggregate grains from the baseline is very modest and near the estimated range of error 
in the grain thermodynamic and optical data of 0.02 dex (5\%) based upon comparisons with other
opacity databases (see F05).
This result implies that including or not including aggregate grains in the computation of 
the mean opacity is not an important decision when computing mean opacity tables.

\subsection{Grain-to-Gas Ratio}

The opacities of F05 were not intended to be used in non-equilibrium conditions.  However,
we can mimic such environments such as conditions of gravitational ``rain-out".  If 
the gas/dust environment is under a gravitational influence with insignificant
radiation pressure, then grains that should appear in the equaton of state 
will settle down into the gravitational potential and ``disappear" from view.  
This will leave behind the apperance of a grain depleted, or a gas rich, environment.
This is best simulated by letting the grains appear in the equation of state, but not
counting their opacities.  Figure~6 shows such simulations.  The baseline amount
from F05 includes 100\% of the grain opacity for each species.  The upper panel of
Fig~6. shows the effects of including 50\%, 10\% and 1\% of the total grain opacity.  

\begin{figure}
%\plotone{opac_comp_grnratio.ps}
\plotone{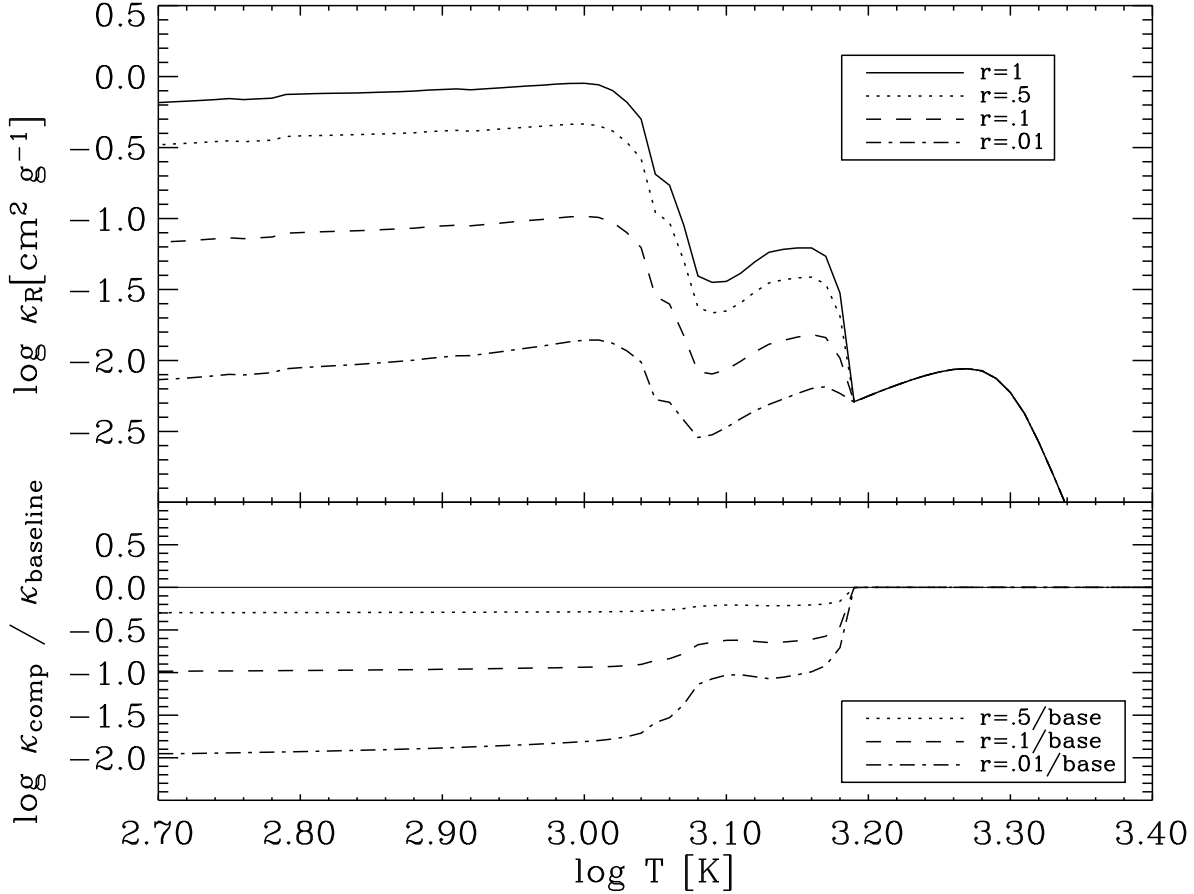}
\caption{Comparisons of total Rosseland mean opacity with varying 
amounts of grain opacity included.  In the upper panel, the solid line is the baseline calcuation 
(100\% of the grain opacity included), the dotted line includes only 50\%, 
the dashed line 10\% and the dash-dot line
only includes 1\% of the grain portion of the opacity.
In the lower panel the differences with the baseline are shown as in Fig.~3 with
the dotted line comparing the 50\% ratio to the baseline, the dashed line
comparing the 10\% ratio, and the dash-dot line the 1\% ratio.
}
\label{Figure 6}
\end{figure}

The effect on the mean opacity is roughly equivalent to the amount of grains removed from 
the grain opacity term.  That is, if the grain opacity is cut in half, then the 
total mean opacity is also roughly half of the baseline value because the grain opacity 
usually completely dominates the total opacity whenever grains are present.  This is
easy to see in the lower panel of Fig.~6.  Note that this effect is not true for
all temperatures however.  At higher temperatures, the differences in the 
opacites (between the baseline and the test cases) are not smooth.  
Just below the grain appearance 
temperature at $\log T \sim 3.2$ there is still a modest molecular opacity present.
If a small amount of grain opacity, say 1\%, is added to the (now) larger molecular opacity, 
the roughly linear scaling does not apply.  As the molecular opacity 
drops at lower temperature, there is a roughly linear relationship between the amount of 
dust included in the monochromatic opacity and the total mean opacity.  This result means
that to a reasonable approximation the opacity tables of F05 could, in principle, be 
modifed to account for the effect of a grainless gas by multiplying by the
desired factor, but only at the coolest temperatures.  It would not be
advisable near the transitions between molecules and grains dominating the opacity.

\section{Discussion}
%### spelling changes
This work implies that caution should be taken when requesting or downloading prefabricated
opacity tables and using them in environments where grain condensation is significant.
%### change in statement to reflect change in Fig. 5.
It is important to include the physical effects that are desired to have accurate opacities, 
most notably changes in the size distribution can greatly affect the Rosseland mean.

Opacity tables for the F05 base set are available for download at our web site at 
{\em http://webs.wichita.edu/physics/opacity}.  Custom opacity tables for non-standard
compositions or for different physical assumptions are
available upon request, as are tables in gas density or gas pressure space instead of $\log~R$.
Note that a full set of opacity tables as outlined in \cite{f05} with (now) 155 values of
hydrogen, X, and metal, Z, abundances take approximately a week of computer time, due to our 
full treatment of the molecular opacity sampling.

\acknowledgments
%### added ack.
We acknowledge the anonymous referee for the suggested references and
other comments that have improved this paper.
Low temperature astrophysics at Wichita State University is supported by 
NSF grant AST-0239590, NASA LTSA grant NAG5-3435 
with matching support from the State of Kansas.  
We also acknowledge the support from the National Science Foundation under 
Grant No. EIA-0216178 and Grant No. EPS-0236913, matching support from the 
State of Kansas and the Wichita State University High Performance Computing Center. 

\newpage
\bibliography{GrainDist}

%
% FIGURE CAPTIONS
% Finally, we have figure captions.  Usually these must be on a separate
% page, although unlike table, it is often permissible to have several
% figure captions on the same page.  We force the page break between
% the reference list and the figure captions.

\end{document}